\documentclass{iopconfser}

\usepackage{bm,color,bbm}

\usepackage{amsmath}
\usepackage{amsfonts}
\usepackage{amssymb}
\usepackage{mathrsfs}
\usepackage{graphicx}
\usepackage{xcolor}

\usepackage{enumitem}

\begin{document}

\title{Bypassing no-go theorems on mixed classical-quantum systems: the (counter)example of hybrid van Hove theory}

\author{Marcel Reginatto$^{1}$, Andr\'{e}s Dar\'{i}o Berm\'{u}dez Manjarres$^{2}$ and Sebastian Ulbricht$^{1,3}$}

\affil{$^1$Physikalisch-Technische Bundesanstalt, Braunschweig, Germany}

\affil{$^2$Universidad Distrital Francisco Jos\'{e} de Caldas, Bogot\'{a}, Colombia}

\affil{$^3$Institut für Mathematische Physik, Technische Universität Braunschweig, Mendelssohnstraße 3, 38106 Braunschweig, Germany}

\email{marcel.reginatto@ptb.de}

\begin{abstract}
Motivated by models of mixed quantum-classical hybrid theories, we consider the representation of classical observables as Hilbert-space operators based on van Hove’s unitary representation of contact transformations. This leads to a reformulation of classical mechanics where the phase space density satisfies the Liouville equation and, more importantly, classical observables are represented by van Hove operators that satisfy a commutation algebra isomorphic to the classical Poisson algebra of functions in phase space. Thus, this Hilbert space approach captures all the essential features of the phase space formulation of classical mechanics in that it reproduces its algebraic structure -- something that is not achieved by other approaches such as the Koopman-von Neumann theory. As an important feature, the definition of appropriate physical states in this formalism imposes certain natural requirements on the phase of the classical wave function, which ensure correct expectation values for the observables. Furthermore, we  discuss how to extend the approach to the \textit{hybrid van Hove theory} of mixed classical-quantum systems and illustrate its application with the example of interacting classical and quantum oscillators. Finally, we focus on the question of how hybrid van Hove theory evades no-go theorems on mixed classical-quantum systems and point to the various assumptions of the no-go theorems that do not apply to our approach. This demonstrates that these no-go theorems are not universally applicable, contrary to their original claims.

\end{abstract}

\section{Introduction}
\label{introduction}

The interplay between classical and quantum mechanics has played an important role in the development of quantum theory, starting with the ``old quantum theory" first formulated by Planck, Einstein, Bohr and Sommerfeld \cite{van1967sources} in which classical and quantum ideas were intermingled, followed by the further developments exemplified by the Copenhagen interpretation, the Wigner-Weyl-Moyal and the de Broglie-Bohm formulations \cite{Heisenberg1930,HILLERY1984121,Holland_1993}, and recent developments which have focused on areas such as decoherence \cite{Joos2003,schlosshauer2007decoherence} and hybrid theories that allow for explicit interactions between classical and quantum systems.

In recent years, a number of no-go theorems regarding hybrid theories have been put forward, some of them arguing that such theories are necessarily inconsistent \cite{PhysRevA.54.3657,Caro_1999,Debendranath_Sahoo_2004,Salcedo_2012}, and others arguing that certain features, like the entanglement of two quantum systems via a classical mediator \cite{Bose2017,MarlettoVedral2017,Galley2022}, are forbidden. These no-go theorems have also been used to argue that the gravitational field would necessarily be quantum or at least non-classical 
if gravitationally mediated entanglement were to be observed.  

In this paper, we use a particular model of interacting classical-quantum systems, Hybrid van Hove theory \cite{Bermudez2024,Reginatto2025,ulbricht2026entanglementquantumsystemsclassical}, to show that the assumptions of the no-go theorems are model dependent and thus not universally applicable. We also discuss which particular assumptions of the no-go theorems do not apply to hybrid van Hove theory.

While both the assumptions and the conclusions of the no-go theorems have been previously challenged \cite{HallReginatto2018,anastopoulos2018commentaspinentanglement,2019JPhCS1275a2038A,Anastopoulos_2021,hall2022commentvindicationentanglementbasedwitnesses,2022FoPh...52..101D,2022AVSQS...4d5601F,dibiagio2026simplereasonclassicalgravity}, the hybrid model discussed in this paper gives new insight into the problem. As we show in the next section, the classical sector of the hybrid theory is a Hilbert space reformulation of classical mechanics which captures its essential features, reproducing in particular the algebraic structure that is characteristic of Hamiltonian mechanics in phase space, something that cannot be achieved by other Hilbert space approaches such as the Koopman-von Neumann theory. Furthermore, in the case of classical-quantum interactions, the theory satisfies essential consistency conditions \cite{Reginatto2025}. 




The paper is structured as follows. We first describe the representation of classical observables in Hilbert space by a set of operators originally introduced by van Hove in his work on the unitary representation of the group of canonical transformations \cite{VanHove1951}. Then, using this representation of classical observables, we work out in detail a formulation of classical mechanics which provides an alternative to the Koopman-von Neumann approach \cite{Bermudez2024,Reginatto2025}.
In the following section we describe the extension to hybrid van Hove theory \cite{Bermudez2024,Reginatto2025} and illustrate its application by the example of interacting classical and quantum oscillators. The last sections are devoted to a discussion on the non-applicability of 
no-go theorems followed by a summary. 

\section{Representation of classical observables in Hilbert space}

Classical mechanics differs from quantum mechanics in important ways: the trajectories of particle are well defined, a property closely related to the absence of an uncertainty principle; the interference effects characteristic of quantum mechanics are absent; and there is a Lie algebra, given in terms of Poisson brackets of phase space functions, which constitute the observables of the theory. A Hilbert space formulation that preserves these and other properties of classical mechanics in phase space must, necessarily, have features that will seem unexpected to those accustomed to the Hilbert space formulation of quantum mechanics. 

That some unusual features are unavoidable becomes clear if we revisit Dirac's original proposal for quantization \cite{C2022}. Dirac
%
listed four rules to be satisfied by any set of quantum operators $\hat{\mathcal{D}}_F$ associated with phase space functions $F(\mathbf{q},\mathbf{p})$ \footnote{There is no universal agreement on the best way of formulating Dirac's rules, so they are sometimes formulated in a slightly different form or with additional requirements \cite{C2022,G2000,G1999}}:
\begin{enumerate}[nolistsep]
\item Identity rule: $\hat{\mathcal{D}}_1 =\hat{1}$
\item Linearity rule: $\hat{\mathcal{D}}_{a F + b G}=a \hat{\mathcal{D}}_{F} + b \hat{\mathcal{D}}_{G}$ 
\item Power rule: $\hat{\mathcal{D}}_{F^n}=(\hat{\mathcal{D}}_F)^n$
\item ``Poisson bracket to commutator'' rule: $[\hat{\mathcal{D}}_F,\hat{\mathcal{D}}_G] = i\hbar \hat{\mathcal{D}}_{\{F,G\}}$
\end{enumerate}
However, Dirac's requirements turned out to be \textit{inconsistent}: no set of operators satisfies \textit{all} the rules simultaneously, as we know from the Groenewold-van Hove theorem \cite{G1999}. Since rules one and two appear to be essential, it is sufficient to focus on rules three and four. Schr\"{o}dinger operators satisfy rule three but not rule four \cite{G1999}. Furthermore, any representation of classical observables in terms of operators preserving the algebraic structure of classical mechanics must satisfy rule four and thus violate rule three. For the case of van Hove operators, introduced below in Eq.~(\ref{vH_operator}), there is no product algebra of observables, meaning that the product of two observables does not result in another observable and therefore rule three is indeed not satisfied. This has far reaching consequences, as we will see in the next sections.  

\subsection{Classical observables from van Hove’s unitary representation of contact transformations}

We now introduce the representation of classical observables in terms of Hilbert space operators. For simplicity, we consider a two-dimensional phase space with coordinates $q$ and $p$. The generalization to more dimensions is straightforward. Given a function $F(q,p)$ of phase space variables, we define the operators  
\begin{equation}
\hat{{\mathcal O}}_F =  F-p\,\partial_p F + i\hbar\left( \partial_q F \,\partial_p - \partial_p F \,\partial_q  \right) \label{vH_operator}\,,
\end{equation}
originally introduced by van Hove \cite{VanHove1951}.  We further consider the operator $\hat{{\mathcal O}}_H$ where $H$ is the Hamilton function $H(q,p)=\frac{p^2}{2m}+V(q)$. The unitary transformation generated by this operator (evolution in time) is described by the equation $i\hbar\frac{\partial\phi}{\partial t}=\hat{{\mathcal O}}_H\phi$. Writing the wave function in polar form, $\phi=\sqrt{\varrho}e^{i\sigma/\hbar}$, we get the pair of uncoupled equations
\begin{eqnarray}  
\frac{\partial\mathcal{\varrho}}{\partial t}+\{\varrho,H\} &=& \quad\frac{d\varrho}{dt} \quad=\quad 0, \label{LE}\\
\varrho \left(  \frac{\partial\sigma}{\partial t}+\{\sigma,H\} \right) &=& \varrho \, \frac{d\sigma}{d t} ~\quad=\quad \varrho L\label{dsigmadt}\,.
\end{eqnarray}
Thus, $\varrho=|\phi|^2$ satisfies the Liouville equation (\ref{LE}), while Eq.~(\ref{dsigmadt}) identifies $\sigma$ with the phase space  \textit{classical action}, given by the time integral of the Lagrange function $L(q,p)=\frac{p^2}{2m}-V(q)$. It is important to take into consideration that this interpretation leads to \textit{consistency conditions} which must be imposed on the phase of the wavefunction: namely that the differential relation $d\sigma = pdq- Hdt$ for the classical action must hold \textit{when it is evaluated over a classical trajectory}. Thus
\begin{equation}\label{constr}
    \frac{\partial \sigma}{\partial q} \simeq p,\quad \frac{\partial \sigma}{\partial p} \simeq 0,\quad \frac{\partial \sigma}{\partial t} \simeq -H, 
\end{equation}
where we introduced the symbol ``$\simeq$'' to indicate that the values of the left- and right-hand sides take equal values when computed over a classical trajectory; i.e., that we do not require the equality to be fulfilled at the level of the phase space functions.

\subsection{From classical densities and action to classical wavefunctions and density operators}
As we saw  in the previous section, the phase space density and the classical action are directly associated with the amplitude and phase of the classical wavefunction. A major difference compared to quantum mechanics is that Eqs.~(\ref{LE})-(\ref{dsigmadt}) for the amplitude and phase are not coupled. This allows us to consider solutions for the amplitude and phase separately.  

\subsubsection{Classical densities in terms of trajectories}
\label{subsubsec:densityAndTrajectories}
We consider a one-dimensional classical system with Hamiltonian $H=\frac{1}{2m}p^2+V(q)$. Let us further introduce the phase space functions
\begin{equation}\label{QPtrajectories}
\tilde Q(q',p',t), \qquad
\tilde P(q',p',t), 
\end{equation}
to denote the classical trajectory that passes through the point $(q',p')$ at time $t=0$.
Any probability density $\rho$ that solves the Liouville equation  (\ref{LE}) can always be represented as a mixture of trajectories,
\begin{equation}\label{rhoTrajectories}
	\varrho(q,p,t) =  \sum_{q',p'} \, w_{q',p'}\delta(q-\tilde Q)\delta(p-\tilde P),
\end{equation} 
with $w_{q',p'} \ge 0$ and $\sum_{q',p'} \, w_{q',p'}=1$. In the case of a continuous distribution, the sum is replaced by an integral, $\sum_{q',p'} w_{q',p'} \rightarrow \int d q' d p' w(q',p')$. In particular, in the case of a single trajectory, we have
\begin{equation}\label{rhoOneTrajectory}
	\varrho'(q,p,t) = \delta(q-\tilde Q(q',p',t))\delta(p-\tilde P(q',p',t)). 
\end{equation} 
In what follows, we will also use the prime notation to denote quantities associated with the trajectory that goes through $(q',p')$ at $t=0$.

\subsubsection{Classical wavefunctions for single trajectories}
We associate the wavefunction
\begin{equation}\label{wftr1}  \phi'=\sqrt{\varrho'}\,e^{i\sigma'/\hbar}
\end{equation}
to the trajectory that passes through $(q',p')$ at $t=0$, with $\varrho'$ given by the delta function\footnote{Although the amplitude of the wave function in Eq.~(\ref{wftr1}) is formally expressed in terms of the square root of the delta function of Eq.~(\ref{rhoOneTrajectory}), this does not present any essential difficulties: all expectation values are well defined and, furthermore, the square root of the delta function can be defined if necessary via a limiting procedure.} of Eq.~(\ref{rhoOneTrajectory}) and the phase by
\begin{equation}\label{sigmaprime}
\sigma':=\eta + H(\tau-\tau'-t),
\end{equation}
where $\eta(q,p)$ is the time-independent solution of Eq.~(\ref{dsigmadt}), $\tau(q,p)$ satisfies $\{\tau(q,p),H(q,p)\}=1$, and $\tau'=\tau(q',p')$. Note that the numerical value of $\tau$ satisfies $\tau \simeq \tau'+t$ when computed over a trajectory. This form of $\sigma'$ is consistent with Eq.~(\ref{constr}), as we now show.

As we already point out in Ref.~\cite{Bermudez2024}, one can evaluate Eq.~(\ref{dsigmadt}) with $\sigma'$, to derive the equality
\begin{equation}
\varrho'(q,p,t) \, \left[ \frac{\partial\sigma'}{\partial t} + \{\sigma',H\}-L\right]\nonumber\\ 
= \varrho'(q,p,t) \left[\frac{\partial\sigma'}{\partial q}\frac{p}{m}-\frac{\partial\sigma'}{\partial p}\frac{\partial V}{\partial q}-\frac{p^{2}}{m} \right] = 0.   
\end{equation}
The delta functions in Eq. (\ref{rhoOneTrajectory}) enforce motion along the trajectories, which leads to $\dot{q} \simeq \frac{p}{m}$ and $\dot{p} \simeq -\frac{\partial V}{\partial q}$ and, therefore,
\begin{equation}
\varrho'(q,p,t) \left[\left(\frac{\partial\sigma'}{\partial q}-p\right)\,\dot{q}+\frac{\partial\sigma'}{\partial p} \dot{p} \right] \simeq 0,  
\end{equation}
which holds true no matter which trajectory or what particular values of $\dot{q}$ and $\dot{p}$ are chosen. It follows that $\varrho'(q,p,t)\left(\frac{\partial\sigma'}{\partial q}-p\right) \simeq 0$ and $\varrho'(q,p,t)\frac{\partial\sigma'}{\partial p} \simeq 0$, such that the first two constraints in Eq.~(\ref{constr}) are satisfied. 
Furthermore, the third constraint $\partial \sigma' / \partial t = -H$ from Eq.~(\ref{constr}) is obviously satisfied, as can be seen from the structure of Eq.~(\ref{sigmaprime}).

Calculating the expectation value
\begin{equation}\label{averageGeneral}
    \langle \hat{\mathcal{O}}_{F} \rangle =
    \int d q d p \, \varrho'\left[F+\left(\frac{\partial \sigma'}{\partial q} - p \right) \frac{\partial F}{\partial p} - \frac{\partial \sigma'}{\partial p} \frac{\partial F}{\partial q} \right],
\end{equation}
we find that
\begin{equation}\label{averageConst}
    \langle \hat{\mathcal{O}}_{F} \rangle \simeq
    \int d q d p \, \varrho' F,
\end{equation}    
by using again the constraints (\ref{constr}) for $\sigma'$. Thus, the expression for the expectation value of a van Hove operator equals the phase space average over the corresponding phase space function, as is required. 

It is important to note some properties of the wavefunctions defined by Eq. (\ref{wftr1}). As we are considering deterministic phase space trajectories, these trajectories cannot cross, which means that for two different trajectories with $(q',p') \neq (q'',p'')$ the corresponding densities $\varrho'$ and $\varrho''$ are defined on two disjunct regions of phase space. Thus, the wave functions are \textit{orthonormal} as they satisfy
\begin{equation}\label{nooverlap}
\int dqdp\, \phi'^\ast(q,p,t)\,\phi''(q,p,t)=0.   
\end{equation}
In addition, since $\tau\simeq\tau'+t$ when $\tau$ is evaluated on a trajectory, the \textit{numerical value} of the wavefunction defined by Eq. (\ref{wftr1}) simplifies to
\begin{equation}\label{valueClassicalWaveFunctionTrajectoriesW} \phi' \simeq  \sqrt{\varrho'}\,e^{i\eta/\hbar},
\end{equation}
with a phase $\eta(q,p)$ that does not depend any longer on the initial conditions $(q',p')$. This, as we show in the next section, allows us to introduce a representation of the state in terms of a wavefunction associated with any density $\varrho$ that solves the Liouville equation, given by
\begin{equation}\label{wf1w} \phi_{\varrho} \simeq \sqrt{\varrho\,}\,e^{i\eta/\hbar}.
\end{equation}
One should keep in mind, however, that the numerical values (\ref{valueClassicalWaveFunctionTrajectoriesW})-(\ref{wf1w}) can be considered only after all manipulations involving van Hove operators acting on states have been carried out. Otherwise expressions 
which rely on the full functional form of the wave function,
like Eq.~(\ref{averageGeneral}), 
would
not evaluate correctly. This is analogous to the handling of \textit{weak equalities} in Dirac's theory of constraints \cite{Dirac:2001:LQM}.

\subsubsection{
Description of general states in terms of a density matrix }
To consider more general states associated with a probability density $\varrho$ satisfying the Liouville equation, we are forced to introduce a description in terms of a density matrix. This prevents the appearance of interference effects, which are excluded in classical systems. The basic idea is the construct arbitrary states from a mixture of phase space trajectories, as opposed to the state described by the wavefunction of Eq. (\ref{wftr1}) which has support on a single trajectory only. The density matrix will then allow us to derive a representation of the state in terms of the wavefunction of Eq. (\ref{wf1w}). 
For that purpose we start with the fact that for \textit{any} Hilbert space with a complete, orthonormal basis $|\gamma_\alpha\rangle$, we can express the most general state of the physical system in terms of a density matrix given by its coefficients $\Pi_{\alpha\beta}$ in that basis. For our case, it is convenient investigate the density matrix in the \textit{coordinate representation} \cite{Landau1981} in phase space and define
\begin{equation}
\pi(q,p;\tilde q,\tilde p) = \sum_{\alpha,\beta}\langle q,p | \gamma_{\alpha} \rangle \Pi_{\alpha\beta} \langle \gamma_{\beta} | \tilde q,\tilde p \rangle = \sum_{\alpha,\beta}\gamma_{\alpha}(q,p) \, \Pi_{\alpha \beta} \, \gamma^*_{\beta}(\tilde q,\tilde p)
\end{equation}
with basis functions $\gamma_{\alpha}(q,p):=\langle q,p | \gamma_{\alpha} \rangle $. In the case in which $\alpha$ and $\beta$ take continuous values, we write instead 
\begin{equation}
\pi(q,p;\tilde q,\tilde p) = \int d\alpha d\beta \; \gamma_{\alpha}(q,p) \, \Pi(\alpha, \beta) \, \gamma^*_{\beta}(\tilde q,\tilde p).
\end{equation}
In particular, we get the density matrix of a single basis state with wavefunction $\gamma_{\lambda}$ by making the diagonal choice $\Pi_{\lambda}(\alpha, \beta)=\delta(\alpha-\beta)\delta(\alpha-\lambda)$ ,
\begin{equation}
\pi_{\lambda}(q,p;\tilde q,\tilde p) = \gamma_{\lambda}(q,p) \, \gamma^*_{\lambda}(\tilde q,\tilde p) = |\gamma_{\lambda}(q,p) \, \gamma^*_{\lambda}(\tilde q,\tilde p)|e^{i(S_{\lambda}(q,p)-S_{\lambda}(\tilde q,\tilde p))}\,,
\end{equation}
where $S_{\lambda}$ is the phase of the basis function. When we evaluate the density associated with $\gamma_{\lambda}$ by setting $\tilde q=q,\tilde p=p$ in $\pi_{\lambda}$, we get $|\gamma_{\lambda}(q,p)|^2=\pi_{\lambda}(q,p;q,p)$.
     
We now use this general result and apply it to our theory, where the complete, orthonormal basis  $\gamma_{\alpha}(q,p)$ is given by the wave functions $\phi_0'(q,p):=\phi'(q,p,t=0)$ defined in Eq. (\ref{wftr1}), where each of them has support on a single trajectory characterized by its initial condition $\lambda\equiv(q',p')$. Then we get
\begin{equation}\label{pi2}
\pi(q,p;\tilde q,\tilde p) = \int dq'dp'dq''dp'' \left[ \phi_0'(q,p) \Pi(q',p',q'',p'') \, \phi_0''^*(\tilde q,\tilde p)\right].
\end{equation}
We further want to specialize to $\Pi=\Pi_\varrho$ with 
\begin{equation}
\Pi_{\varrho}(q',p',q'',p'') = \sqrt{\varrho(q',p')\varrho(q'',p'')} \delta(q'-q'')\delta(p'-p''),
\end{equation} 
where $\varrho$ is any density at time $t=0$.

The resulting density matrix in coordinate representation $\pi=\pi_{\varrho}$ can be used to calculate expectation values for all operators $\hat{\mathcal{O}}_{F}$.
However, if we only care about the \textit{numerical value} of $\pi_{\varrho}$, we can use Eq. (\ref{valueClassicalWaveFunctionTrajectoriesW}) instead of Eq. (\ref{wftr1}) in Eq. (\ref{pi2}), as either of these sets of wavefunctions provides a complete, orthonormal basis, the main difference being that the set of Eq. (\ref{valueClassicalWaveFunctionTrajectoriesW}) does not satisfy the constraint of Eq. (\ref{constr}). This change leads to 
\begin{eqnarray}
\pi_{\varrho}(q,p;\tilde q, \tilde p) &\simeq& \int dq'dp'dq''dp'' \left\{|\phi'(q,p)| \;\Pi_{\varrho}(q',p',q'',p'') \; |\phi''(\tilde q,\tilde p)|\right\}e^{i (\eta-\eta')/\hbar} 
\end{eqnarray}
We now associate a wavefunction $\phi_{\varrho}(q,p)$ to the numerical value of $\pi_{\varrho}$. It is clear from the equation above that its phase will be $\eta(q,p)/\hbar$. To get the \textit{density} of the wavefunction associated with this density operator, we set  $\tilde q=q,\tilde p=p$ in $\pi_{\varrho}$, which leads to
\begin{equation}
\pi_{\varrho}(q,p;q,p) \simeq \varrho(q,p)\,. 
\end{equation}
Thus, we define the wavefunction by $\phi_{\varrho}(q,p) \simeq \sqrt{\varrho(q,p)}e^{i\eta(q,p)/\hbar}$. We can generalize this result to a time-dependent wavefunction $\phi_\varrho(q,p,t)$ by simply requiring that the density solves the Liouville equation; this is the wavefunction of Eq.~(\ref{wf1w}). The converse is also true: For any $\phi_{\varrho}(q,p,t)$ that is of the form given in Eq. (\ref{wf1w}), there is a corresponding density operator in the coordinate representation which is of the form $\pi_{\varrho}(q,p;\tilde q,\tilde p)$.

\subsection{Heisenberg representation}
\label{subsec:Heisenberg}

In the the Heisenberg representation, the operators $\hat{\mathcal{O}}_{F}(t)$ are time-dependent with equation of motion 
\begin{equation}
 i\hbar \frac{\partial}{\partial t}{\hat{\mathcal{O}}}_{F}(t) = [\hat{\mathcal{O}}_{F}(t),\hat{\mathcal{O}}_{H}] \label{eqn:Heißenberg_equation}
\end{equation}
and the wavefunctions are time-independent. Given that an operator $\hat{\mathcal{O}}_{F}(t)$ at time $t=0$ equals the time-independent van Hove operator from Eq.~(\ref{vH_operator}), such that $\hat{\mathcal{O}}_{F}(0)=\hat{\mathcal{O}}_{F}$, the time-dependent operator can be expressed in the usual way, $\hat{\mathcal{O}}_{F}(t) = e^{i\hat{\mathcal{O}}_H t/\hbar}\, \hat{\mathcal{O}}_{F}\, e^{-i\hat{\mathcal{O}}_H t/\hbar}$. \footnote{Note that $e^{i\hat{\mathcal{O}}_H t/\hbar}\, \hat{\mathcal{O}}_{F}\, e^{-i\hat{\mathcal{O}}_H t/\hbar}=\hat{\mathcal{O}}_{F}+i [\hat{\mathcal{O}}_{H},\hat{\mathcal{O}}_{F}] t/\hbar-  [\hat{\mathcal{O}}_{H},[\hat{\mathcal{O}}_{H},\hat{\mathcal{O}}_{F}]]t^2/2\hbar^2+\dots$ contains only commutators and no operator product. 
}
Concurrently, the wavefunction in this representation becomes time-independent and is given by Eqs.~(\ref{rhoOneTrajectory})-(\ref{sigmaprime}) evaluated at $t=0$. 

We check that the phase of the time-independent wavefunction satisfies the first two constraints of Eq.~(\ref{constr}) so that the expectation values in Eq.~(\ref{averageGeneral}) reduce to Eq.~(\ref{averageConst}). To see that this is indeed the case, consider the time-dependent wavefunction, Eq.~(\ref{sigmaprime}), for a time $\Delta t$ that is infinitesimally close to $t=0$. Then the motion along the trajectory can be approximated by motion along the tangent to the trajectory at $t=0$; i.e., free particle motion. We can therefore set $H \approx p^2/2m$, $\tau \approx mq/p$, $\tau' \approx mq'/p'$ and $\eta \approx qp/2$ in Eq.~(\ref{sigmaprime}) and
\begin{equation}
    \sigma' = \eta + H(\tau-\tau'-\Delta t) \approx \frac{q p}{2} + \frac{p^2}{2m}\left(\frac{mq}{p}-\frac{mq'}{p'}-\Delta t\right), 
\end{equation}
which leads to $\frac{\partial \sigma'}{\partial q} \approx p$ and $\frac{\partial \sigma'}{\partial p} \approx 0$, as required, since the $\tau - \tau' \simeq \Delta t$ when evaluated over a trajectory. These expressions become exact as $\Delta t \rightarrow 0$. Thus Eq.~(\ref{averageConst}) also holds in the Heisenberg representation. Alternatively, since $\varrho'=\delta(q-q')\delta(p-p')$ at time $t=0$, the first two constraints of Eq.~(\ref{constr}) can be satisfied by setting $\sigma'=qp'$ in Eq. (\ref{sigmaprime}), since then $\frac{\partial \sigma'}{\partial q} = p'$ and $\frac{\partial \sigma'}{\partial p} = 0$. This choice is convenient for calculations. 

More general states with arbitrary $\varrho$ are handled by introducing the corresponding density matrix, as we did in the previous section.

\section{Example: Interacting oscillators in Hybrid van Hove theory} \label{sec:oscillators}
In hybrid van Hove theory \cite{Bermudez2024,Reginatto2025}, the observables of the classical particle are represented by van Hove operators, the observables of the quantum particle by Schr\"{o}dinger operators, and the Hamiltonian includes a hybrid interaction term that is the product of a classical and a quantum operator.  In previous publications, we have described how hybrid van Hove theorey is related to other hybrid formulations \cite{Bermudez2024}, looked at the consistency of the approach \cite{Reginatto2025},  and have shown that entanglement can be generated via a classical mediator \cite{ulbricht2026entanglementquantumsystemsclassical}. In this section we discuss a simple example to illustrate the application of the theory: a hybrid system of two particles where a classical oscillator interacts with a quantum oscillator.

Given a phase space function $F(q,p)$, the corresponding classical and quantum observables are represented by the van Hove operator $\hat{\mathcal{O}}_{F(q,p)}$ and the Schr\"{o}dinger operator $\hat{F}(q,p)$, respectively. 
Thus, in this formulation, we have a clear algebraic distinction between the set of classical observables and the set of quantum observables: the operators $\hat{\mathcal{O}}_{F(q,p)}$ satisfy \textit{different} commutator relations and, therefore, a \textit{different} Lie algebra than the operators $\hat{F}(q,p)$. This distinction between classical and quantum systems is preserved in the case of hybrid systems.
We consider first the purely classical Hamiltonian 
\begin{equation}\label{intOsc}
	H = \left[\left(\frac{p^2}{2m} + \frac{m\omega^2q^2}{2}\right) + \left(\frac{y^2}{2m} + \frac{m\omega^2x^2}{2}\right)\right] + \lambda qx,
\end{equation}
with phase space coordinates $q,p$ for the first, and $x,y$ for the second particle. We find the solution of the classical system by going to normal modes, using an algebraic approach. When we carry out the linear canonical transformation
\begin{equation}
    \qquad X^{\pm} = \sqrt{\frac{1}{2}}(q \pm x), \qquad P^{\pm} = \sqrt{\frac{1}{2}}(p \pm y), 
\end{equation}
the Hamiltonian takes the form
\begin{equation}\label{HNM}
    H = \frac{(P^{+})^2}{2m}+\frac{(P^{-})^2}{2m}+\frac{m(\Omega^{+})^2(X^{+})^2}{2}+\frac{m(\Omega^{-})^2(X^{-})^2}{2},
\end{equation}
where $\Omega^{\pm}:=(\omega^2 \pm {\lambda}/{m})^{1/2}$. More generally, we may consider the set of, generally time dependent, observables 
\begin{equation}\label{closedLieAlgObs}
I = 1, \qquad X^{\pm}, \qquad P^{\pm}, \qquad (X^{\pm})^2/2, \qquad (P^{\pm})^2/2, \qquad X^{\pm}P^{\pm},
\end{equation}
which are polynomials of up to second order and  form a closed set under Poisson brackets, with the non-zero Poisson brackets given by
\begin{align}\label{closedLieAlg}
\{X^{\pm},P^{\pm}\} &= I, & \{X^{\pm},(P^{\pm})^2/2\} &= P^{\pm}, & \{X^{\pm},X^{\pm}P^{\pm}\} &= X^{\pm}, \nonumber\\
\{P^{\pm},(X^{\pm})^2/2\} &= - X^{\pm}, &  \{P^{\pm},X^{\pm}P^{\pm}\} &= -P^{\pm},  & \{(X^{\pm})^2/2,(P^{\pm})^2/2\} &=  X^{\pm}P^{\pm}, \nonumber\\
\{(X^{\pm})^2/2, X^{\pm} P^{\pm} \} &= (X^{\pm})^2, & 
\{(P^{\pm})^2/2,X^{\pm}P^{\pm}\} &=  -(P^{\pm})^2. 
\end{align}
As the Hamiltonian of Eq. (\ref{HNM}) is written in terms of the observables of Eq. (\ref{closedLieAlgObs}), is is straightforward to write the equations of motion for the normal mode coordinates $X^{\pm}$,
\begin{equation}\label{closedLieAlgMotion}
\frac{d^2}{dt^2}X^{\pm} = \{X^{\pm},\{X^{\pm},H\}\}=-(\Omega^{\pm})^2 X^{\pm},
\end{equation}
with solution
\begin{equation}
    X^{\pm}(t)=X^{\pm}(0)\cos(\Omega^{\pm} t)+\frac{1}{m\Omega^{\pm}}P^{\pm}(0)\sin(\Omega^{\pm} t).
\end{equation}
for given initial conditions $X^\pm(0)$ and $P^\pm(0)$.
It is worth looking in some detail at the solution of the interacting classical oscillators described by the Hamiltonian of Eq. (\ref{HNM}) because the algebraic approach used here, and in particular the Lie algebra of Eq. (\ref{closedLieAlg}), carries over to the hybrid case \textit{without any modifications}. This is a consequence of the following properties of operators: (i) the commutator algebra of the van Hove operators is isomorphic to the Poisson algebra of functions in phase space \cite{VanHove1951}, (ii) the operators representing classical and quantum observables satisfy the \textit{locality} property
\begin{equation}
    [\hat{\mathcal{O}}_{F(q,p)},\hat{G}(q,p)]=0,
\end{equation} 
as one can determine by direct calculation, and (iii) for polynomials of up to second order, the operators representing quantum observables have a commutator algebra that agree with the Poisson algebra of the corresponding functions in phase space\footnote{ This equivalence does not hold for polynomials of higher order, as is well known from the Groenewold-van Hove theorem.} \cite{VanHove1951}. 

To describe hybrid interacting oscillators, we take the Hamiltonian of Eq.~(\ref{intOsc}) and replace the functions of $q,p$ in the first parenthesis by van Hove operators, the functions of $x,y$ in the second parenthesis by Schr\"{o}dinger operators, and the interaction term by its corresponding hybrid operator,
\begin{equation}\label{intHOsc}
	\hat{\Theta}_{H} = \left[\left(\frac{\hat{\mathcal{O}}_{p^2}}{2m} + \frac{m\omega^2\hat{\mathcal{O}}_{q^2}}{2}\right) + \left(\frac{\hat{y}^2}{2m} + \frac{m\omega^2\hat{x}^2}{2}\right)\right] + \lambda \hat{\mathcal{O}}_{q}\hat{x},
\end{equation}
where we introduced the notation $\hat{\Theta}$ for hybrid operators. The hybrid operator $\lambda \hat{\mathcal{O}}_{q}\hat{x}$ that corresponds to the classical interaction term $\lambda qx$ is uniquely determined, as shown in Appendix \ref{App:InterTerm}. Analogous to the classical case discussed above, we introduce the hybrid operators
\begin{equation}
    \hat{\Theta}_{X^{\pm}} := \sqrt{\frac{1}{2}}(\hat{\mathcal{O}}_q \pm \hat{x}), \qquad \hat{\Theta}_{P^{\pm}} := \sqrt{\frac{1}{2}}(\hat{\mathcal{O}}_p \pm \hat{y}). 
\end{equation}
Applying the Heisenberg representation, introduced in Sec.~\ref{subsec:Heisenberg}, to the hybrid operator $\hat{\Theta}_{X^\pm}(t) = e^{i\hat{\Theta}_H t/\hbar}\, \hat{\Theta}_{X^\pm}\, e^{-i\hat{\Theta}_H t/\hbar}$, the equation of motion for the hybrid normal modes takes the form
\begin{equation}
\frac{\partial^2}{\partial t^2} \hat{\Theta}_{X^\pm}(t) =(\Omega^{\pm})^2\hat{\Theta}_{X^\pm}(t)\,,
\end{equation} 
with the solution
\begin{equation}
	\hat{\Theta}_{X^{\pm}}(t) = \hat{\Theta}_{X^{\pm}}\cos(\Omega^{\pm} t)+\frac{1}{m\Omega^{\pm}}\hat{\Theta}_{P^{\pm}}\sin(\Omega^{\pm} t).
\end{equation}

Peres and Terno have argued that purely classical and quantum interacting oscillators of the type discussed in this section share certain algebraic properties which ought to be preserved by any mixed classical-quantum theory that describes the corresponding hybrid system. They considered this requirement a ``definitive benchmark [...] for an acceptable quantum-classical hybrid system'' \cite{PT2001}. As we have explicitly shown with the example in this section, hybrid van Hove theory satisfies this benchmark. However, while this is true of a few other hybrid models (e.g., ensembles on configuration space \cite{HR2016} and Elze's model \cite{PhysRevA.85.052109}), many hybrid theories do not incorporate the algebraic structure of observables that is familiar to us from classical and quantum mechanics, making it impossible to give a straightforward algebraic solution to this simple example (e.g., Sudarsan's model based on Koopman-von Neumann \cite{Sudarshan1976} and models with commuting classical operators originating in quantum information theory \cite{Bose2017,MarlettoVedral2017}, among others).

\section{Why current no-go theorems on hybrid models do not apply to hybrid van Hove theory} \label{sec:no-go}

No-go theorems have claimed that consistent hybrid models are not possible \cite{PhysRevA.54.3657,Caro_1999,Debendranath_Sahoo_2004,Salcedo_2012} or that they cannot allow for certain properties, like entanglement via a classical mediator \cite{Bose2017,MarlettoVedral2017,Galley2022}. While the no-go theorems are mathematically correct, their relevance and/or universality have been questioned, as we have already pointed out in Section \ref{introduction}: the no-go theorems are typically based on assumptions that, as we discuss below, are too restrictive, and while some hybrid models may satisfy them, they are not universally applicable. We now illustrate some of the shortcomings of no-go theorems with the example of hybrid van Hove theory. 

\subsection{The issue of ``classicality'' and its meaning in hybrid van Hove theory}
While there is no strict definition of ``classicality'', the term typically refers to those \textit{defining features} that distinguish a classical system from a quantum system. The way classicality is interpreted varies among the different no-go theorems, as they emphasize different aspects of a classical system: the definition that emerges from quantum information theory \cite{MarlettoVedral2017} differs dramatically from the definition adopted for generalized probability theories (GPT) \cite{Galley2022}, and these in turn do not correspond with classicality as defined for example in the hybrid theories of ensembles on configuration space \cite{HR2016} or those based on hybrid extensions of the de Broglie-Bohm theory \cite{Doner2022}. Therefore, to understand how hybrid van Hove theory bypasses the no-go theorems, it becomes crucial to explain the notion of classicality assumed in this hybrid model.

For the sake of simplicity, here we illustrate these points using the example of
a hybrid system of a classical particle interacting with a quantum particle. In hybrid van Hove, the following ``classicality'' properties hold for the classical sector:

\begin{enumerate}
    \item the particle is assigned (continuous) phase space coordinates $q$ and $p$,
    \item each classical observable $F(q,p)$ in phase space is represented in Hilbert space by its corresponding van Hove operator $\hat{{\mathcal O}}_F$, 
    \item  the commutator algebra of the set of observables $\hat{{\mathcal O}}_F$ is isomorphic to the algebra of Poisson brackets for the set of phase space observables $F(q,p)$, implying that classical observables $\hat{{\mathcal O}}_F$  are \textit{not} represented by commuting observables,
    \item while the commutator of two classical observables is again a classical observable, there is no product algebra for the operators  $\hat{{\mathcal O}}_F$, which means that the product of two such operators is not a classical observable, 
    \item for all classical operators $\hat{{\mathcal O}}_F$ and quantum operators $\hat{G}$ we have $[\hat{{\mathcal O}}_F,\hat{G}]=0$, a locality condition implying that a unitary transformation that only affects the classical (quantum) particle cannot generate a transformation of quantum (classical) observables,
    \item the statistical description of the classical particle at any given time requires knowledge of the evolution of the full hybrid state as determined by the hybrid Hamiltonian operator; in particular, it is not possible to introduce a single ``reduced'' classical Hamiltonian operator operating on the classical sector to describe how the statistics of the classical particle evolves \cite{ulbricht2026entanglementquantumsystemsclassical}.
\end{enumerate}   

Thus, in hybrid van Hove theory, the classical sector preserves all the algebraic properties of classical mechanics as a consequence of using the operators of van Hove, which provide a unitary representation of the Lie algebra of the generators of canonical transformations. This means that the requirement of Peres and Terno \cite{PT2001} are satisfied, as well as several requirements of Salcedo \cite{Salcedo_2012}. However, it also means that the assumptions of various no-go theorems do not apply to hybrid van Hove theory, as we now explain.

\subsection{On the limitations of no-go theorems on hybrid systems} 
The literature on hybrid models is extensive and it is impossible to address every single no-go theorem. This selection of the no-go theorems that have received the most attention is already sufficient to indicate how the claim of universality validity often made for these theorems is not valid and to understand how hybrid van Hove theory evades these no-go theorems.

\subsubsection{The no-go theorems of Marletto and Vedral} The hybrid models considered are in this class of no-go theorems based on notions of quantum information theory where the classical observables are represented by \textit{commuting operators} \cite{MarlettoVedral2017}, which contradicts properties 2, 3 and 4.
\subsubsection{The no-go theorem of Bose et al} It is assumed that the classical mediator acts as a classical communication channel within the framework of \textit{local operations and classical communication} (LOCC) \cite{Bose2017}, which is a notion of classicality that is very far from the notion of classicality in hybrid van Hove theory. Furthermore, this kind of classical communication between the quantum systems can always be represented by commuting operators in a sufficiently large Hilbert space that includes the two quantum systems and the mediator \cite{HallReginatto2018}, which contradicts properties 1, 2, 3 and 4.
\subsubsection{No-go theorems assuming the hybrid systems is a \textit{generalized probabilistic theory}} In a GPT, it is assumed that the classical system is a discrete simplex \cite{Galley2022}, with no further structure like an algebra, which contradicts properties 2 and 3. Furthermore, in a GPT there are assumptions about the equations of motion of classical subsystems \cite{Galley2022} that do not apply to hybrid van Hove theory, in that the evolution of the classical subsystem does not necessarily follow a reduced description given in terms of purely classical equations, contradicting property 6.

\subsubsection{No-go theorems assuming the Koopman-von Neumann approach} The representation of the classical sector in terms of the Koopman-von Neumann theory, which is considered for example in papers by Terno \cite{T2006} and Marletto et al \cite{marletto2025entanglementmediatedkoopmaniansystem} contradicts properties 2, 3 and 4.

\subsubsection{No-go theorems requiring a product algebra} Some no-go theorems \cite{PhysRevA.54.3657,Caro_1999,Debendranath_Sahoo_2004} assume that the set of observables can be extended to form a product algebra, which contradicts property 4.

\section{Summary}
We have reviewed in detail a reformulation of classical mechanics in Hilbert space that makes use of the operators introduced by van Hove \cite{VanHove1951} to represent classical observables. This correspondence leads to an isomorphism between the Poisson bracket algebra of functions in phase space and the commutator algebra of the corresponding van Hove operators. The definition of the classical states requires some care: states that have support on classical trajectories can be represented by orthonormal wavefunctions that satisfy a constraint on their phase, while most general densities are represented by density matrices. In this formulation, the Born rule implies that the density associated with the wavefunction satisfies the Liouville equation, the phase of the wave function is interpreted as the classical action, and all expectation values equal the average values of the corresponding phase space functions with respect to the phase space density.  

The classical theory in Hilbert space can be extended in a natural way to hybrid van Hove theory, a model of interacting classical and quantum systems where the classical observables are represented by van Hove operators and the quantum observables by the usual operators of a quantum system. 

Hybrid van Hove theory satisfies consistency requirements \cite{Reginatto2025} and allows for correlations between classical and quantum systems that can lead to the entanglement of quantum systems via a classical mediator \cite{ulbricht2026entanglementquantumsystemsclassical}. This is clearly in contradiction with a number of no-go theorems which have claimed to be universally applicable and have received much attention in the past few years. As we discussed in detail in Sec.~\ref{sec:no-go}, our theory evades these no-go theorems because of the limited applicability of these theorems. The assumptions of the no-go theorems are too restrictive, as we show explicitly by pointing to those assumptions that are not made in hybrid van Hove theory, or clearly contradict the assumption of hybrid van Hove theory. 

\section*{Acknowledgment}
SU acknowledges funding by the Deutsche Forschungsgemeinschaft (DFG, German Research Foundation) under Germany’s Excellence Strategy—EXC 2123 QuantumFrontiers—390837967. 

\appendix

\section{Interaction terms}
\label{App:InterTerm}

Here, we have a brief look on the uniqueness of hybrid interaction Hamiltonians in hybrid van Hove theory.
Consider a classical Hamiltonian in phase space of the form
\begin{equation}\label{A}
H(q,p,x,y) = H_1(q,p) + H_2(x,y) + I(q,p,x,y),
\end{equation} 
which describes two particles (one with coordinates $q,p$ and Hamiltonian $H_1$ and the other one with cooordinates $x,y$ and Hamiltonian $H_2$) coupled via an interaction term of the form
\begin{equation}\label{A}
I(q,p,x,y) = \sum_k b_k(q,p)c_k(x,y)
\end{equation} 
for some functions $b_k$ and $c_k$. This form is general enough to include a large class of Hamiltonians, for example
\begin{equation}\label{int2part}
	H = H_1(q,p) + H_2(x,y) +\lambda qx = \left[\left(\frac{p^2}{2m} + V_1(q)\right) + \left(\frac{y^2}{2m} + V_2(x)\right)\right] + \lambda qx,
\end{equation}
that is used in the example of two coupled oscillators from Sec.~\ref{sec:oscillators},
as well as the Hamiltonians that are used to describe \textit{von Neumann measurements} where the interaction term is proportional to e.g. products of position and momentum like $qy$ or $xp$ \cite{vonNeumann1955}. 

We now introduce the corresponding \textit{hybrid system} where the particle with coordinates $q,p$ remains classical while the particle with coordinates $x,y$ is quantized. We will show that $H$ is mapped to a \textbf{unique hybrid operator}, which we will denote by $\hat{\Theta}_H$. To see that the operator $\hat{\Theta}_H$ is indeed unique, consider the two possible  ways of defining such an operator:
\begin{enumerate}[nolistsep]
	\item First promote functions with coordinates $x,y$ to Schr\"{o}dinger operators while treating what remains as parameters, then promote the functions with coordinates $p,q$ that remain to van Hove operators,
	\begin{eqnarray}{}
		H = H_1 + H_2 + \sum_kb_kc_k &\xrightarrow[\text{Sch.}]{}& H_1 + \hat{H}_2 + \sum_kb_k \hat{c}_k \nonumber\\
		&\xrightarrow[\text{v H.}]{}& \hat{\mathcal{O}}_{H_1} + \hat{H}_2 + \sum_k \hat{\mathcal{O}}_{b_k} \hat{c}_k =: \hat{\Theta}_H^{(1st~ way)}.\qquad 
	\end{eqnarray} 
	\item First promote functions with coordinates $q,p$ to van Hove operators while treating what remains as parameters, then promote the functions with coordinates $x,y$ that remain to Schr\"{o}dinger operators,
    \begin{eqnarray}
        H = H_1 + H_2 + \sum_kb_kc_k 
        &\xrightarrow[\text{v H.}]{}& \hat{\mathcal{O}}_{\left(H_1+\sum_k b_k c_k\right)} + H_2 \quad = \hat{\mathcal{O}}_{H_1} + \sum_k \hat{\mathcal{O}}_{b_k} c_k  + H_2 \nonumber\\ 
		&\xrightarrow[\text{Sch.}]{}& \hat{\mathcal{O}}_{H_1} + \sum_k \hat{\mathcal{O}}_{b_k} \hat{c}_k  + \hat{H}_2 =: \hat{\Theta}_H^{(2nd~ way)}, 
    \end{eqnarray}
	where we used the identity $\hat{\mathcal{O}}_{(F+G)}=\hat{\mathcal{O}}_{F}+\hat{\mathcal{O}}_{G}$, where $F$ and $G$ are arbitrary functions.
\end{enumerate}
Thus both ways of defining $\hat{\Theta}_H$ lead to the same result, 
	$\hat{\Theta}_H = \hat{\mathcal{O}}_{H_1} + \hat{H}_2 + \sum_k \hat{\mathcal{O}}_{b_k} \hat{c}_k,$
which shows that there is a unique hybrid operator associated with $I$ as given by Eq. (\ref{A}).

\bibliographystyle{iopart-num}

\bibliography{literature}

\end{document}